\documentclass{emulateapj}

\shorttitle{ASTEROSEISMIC DIAGRAMS}
\shortauthors{WHITE ET AL.}
\submitted{Accepted by ApJ}

\begin{document}

\title{Calculating asteroseismic diagrams for solar-like oscillations}

\author{Timothy R. White\altaffilmark{1,2}, Timothy R. Bedding\altaffilmark{1}, Dennis
Stello\altaffilmark{1}, J{\o}rgen Christensen-Dalsgaard\altaffilmark{3}, Daniel
Huber\altaffilmark{1}, and Hans Kjeldsen\altaffilmark{3}}
\altaffiltext{1}{Sydney Institute for Astronomy (SIfA), School of Physics, University
of Sydney, NSW 2006, Australia; t.white@physics.usyd.edu.au}
\altaffiltext{2}{Australian Astronomical Observatory, PO Box 296, Epping NSW 1710,
Australia}
\altaffiltext{3}{Danish AsteroSeismology Centre (DASC), Department of Physics and
Astronomy, Aarhus University, DK-8000 Aarhus C, Denmark}

\begin{abstract}
\noindent With the success of the {\it Kepler} and {\it CoRoT} missions, the number of stars with detected
solar-like oscillations has increased by several orders of magnitude, for the first time we are able to
perform large-scale ensemble asteroseismology of these stars. In preparation for this golden age of asteroseismology we
have computed expected values of various asteroseismic observables from models of varying mass and metallicity. The
relationships between these asteroseismic observables, such as the separations between mode frequencies, are able to
significantly constrain estimates of the ages and masses of these stars. We investigate the scaling relation between the
large frequency separation, $\Delta\nu$, and mean stellar density. Furthermore we present model evolutionary tracks for
several asteroseismic diagrams. We have extended the so-called C-D~diagram beyond the main sequence to the subgiants and
the red-giant branch. We also consider another asteroseismic diagram, the $\epsilon$~diagram, which is more sensitive to
variations in stellar properties at the subgiant stages and can aid in determining the correct mode identification. The
recent discovery of gravity-mode period spacings in red giants forms the basis for a third asteroseismic diagram. We
compare the evolutionary model tracks in these asteroseismic diagrams with results from pre-{\it Kepler} studies of
solar-like oscillations, and early results from {\it Kepler}. \end{abstract}


\section{Introduction}
Asteroseismology promises to expand our knowledge of the stars through the study of
their oscillations. This promise has driven efforts to measure oscillations in
solar-type stars with ground-based observations, but the requirement for precise
measurements (at the level of m\,s$^{-1}$ in radial velocity) that are well-sampled
over a long period of time have limited the number of detections to only a handful of
stars \citep[see][for recent reviews]{Aerts08, Bedding11b}. 

Space-based missions are ideal to ensure continuous data sets and the {\it CoRoT} satellite has
measured oscillations in several stars \citep[e.g.][]{Michel08}. The {\it Kepler
Mission} is set to revolutionize the study of oscillations in main-sequence and subgiant
stars by increasing the number of stars with high-quality observations by more than two orders of magnitude 
\citep{Gilliland10, Chaplin11a}. 

With the large number of stars observed by {\it Kepler} it becomes possible to perform ensemble asteroseismology of
stars with solar-like oscillations. This includes constructing asteroseismic diagrams, in which different measurements
of the oscillation spectrum are plotted against each other, revealing features that are dependent upon the stellar
structure.

For acoustic modes of high radial order, $n$, and low angular degree, $l$, frequencies are well-approximated by the
asymptotic relation \citep{Vandakurov67, Tassoul80, Gough86}:
\begin{equation}
\nu_{n,l}\approx\Delta\nu\left(n+{l \over 2}+\epsilon\right)-\delta\nu_{0l}.\label{asymp}
\end{equation}
Here, $\Delta\nu$ is the so-called large separation between modes of the same $l$ and consecutive $n$, while
$\delta\nu_{0l}$ is the small separation between modes of different $l$, and $\epsilon$ is a dimensionless offset. To a
good approximation, $\Delta\nu$ is proportional to the square root of the mean density of the star \citep{Ulrich86} and in Section
\ref{scale} we investigate the validity of this approximation. The small separations, $\delta\nu_{0l}$, are sensitive to
the structure of the core and  hence to the age of the star, at least on the main sequence. These somewhat orthogonal
dependencies leads to their use in the so-called C-D~diagram, in which the large and small separations are plotted against
each other \citep{C-D84}. Calculating the C-D~diagram is one of the main aims of this paper.

Previous studies of the C-D~diagram and its variations have determined the expected evolution of stars with varying mass
and metallicity \citep{Ulrich86,Gough87,C-D88}, and assessed the feasibility of applying the diagram to real data
\citep{Monteiro02,OtiFloranes05,Mazumdar05,Gai09}. However, none of these studies followed the evolution beyond the
end of the main sequence. Recently, \citet{Montalban10} computed the theoretical spectrum of solar-like oscillations in
red-giant stars, finding that the small separation $\delta\nu_{02}$ depends almost linearly on $\Delta\nu$, in agreement
with the red-giant results from {\it Kepler} \citep{Bedding10c, Huber10}. In Section \ref{CD} we bridge the gap,
extending the C-D~diagram beyond main-sequence stars to the subgiants and up towards the tip of the red-giant branch.

A complication with the C-D~diagram for subgiants and red-giant stars arises from mode bumping. As stars evolve, the convective
envelope expands and the acoustic oscillation modes ($p$ modes) decrease in frequency. At the same time, $g$-mode
oscillations that exist in the core of the star increase in frequency as the core becomes more centrally condensed.
Eventually, $p$- and $g$-mode frequencies overlap, resulting in oscillation modes that have a mixed character, behaving
like $g$ modes in the core and $p$ modes in the envelope. The frequencies of these modes are shifted as they undergo
avoided crossings \citep{Osaki75, Aizenman77}, which leads to significant deviations from the asymptotic relation,
equation~(\ref{asymp}). This so-called mode bumping only affects non-radial modes, particularly $l$=1 but also $l$=2,
and so it complicates the measurement of the small separations. Nevertheless,
as we show, it is still possible to measure average separations that can be plotted in the C-D~diagram.

In this paper we also discuss an asteroseismic diagram that uses the quantity $\epsilon$ (Section \ref{Epsilon}).
Despite being investigated by \citet{C-D84}, this dimensionless phase offset has since been largely overlooked 
for its diagnostic potential. Recently, \citet{Bedding10b} suggested that it could be useful in distinguishing odd 
and even modes when their identifications are ambiguous due to short mode lifetimes 
\citep[see, e.g.,][]{Appourchaux08,Benomar09,Bedding10}. Using {\it Kepler} data, \citet{Huber10} have found that 
$\epsilon$ and $\Delta\nu$ are related in red giants, implying that, like $\Delta\nu$, $\epsilon$ is a function of 
fundamental parameters. A similar analysis was done for {\it CoRoT} data by \citet{Mosser11}. We discuss the use
of $\epsilon$ for mode identification in Section \ref{modeID}.

Finally, we discuss an asteroseismic diagram for red-giant stars. A recent breakthrough has been made with the discovery
of sequences of mixed modes in {\it Kepler} red giants \citep{Beck11}. Because these mixed modes exhibit $g$-mode
behavior in the core of the star, they are particularly sensitive to the core structure. Subsequently,
\citet{Bedding11} used the observed period spacings of these mixed modes, $\Delta P_\mathrm{obs}$, to distinguish between
red giants that are burning helium in their core and those that are still only burning hydrogen in a shell. 
\citet{Mosser11b} have found similar results in {\it CoRoT} red giants. In Section \ref{Gperiod} we present the 
expected evolution of $\Delta P_\mathrm{obs}$ with $\Delta\nu$ in an asteroseismic diagram for red giant stars. 

\section{Measuring Asteroseismic Parameters from Models}\label{Measure}

\begin{figure}
\epsscale{1.2}
\plotone{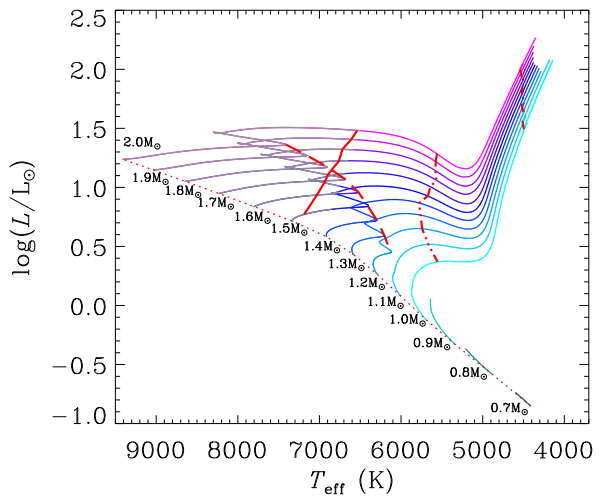}
\caption{H-R diagram of models with near-solar metallicity ($Z_0=0.017$) of masses from 0.7 (green) to
$2.0\,\mathrm{M}_\odot$ (magenta). The section of tracks that are gray are hotter than the approximate cool edge of the classical
instability strip \citep{Saio98}. The zero-age main sequence is indicated by the dotted red line. Other red lines
(solid, long dashed, dot-dashed and short dashed) relate to features in Figure \ref{fig4}.}\label{fig1}
\end{figure}

A grid of 51000 stellar models was calculated from the ZAMS to almost the tip of the red-giant branch using \textsc{ASTEC}
\citep{C-D08a} with the EFF equation of state \citep{Eggleton73}. We used the opacity tables of \citet{RogersIglesias95}
and \citet{Kurucz91} for $T<10^4$\,K, with the solar mixture of \citet{GrevesseNoels93}. Rotation, overshooting and diffusion
were not included. The grid was created with fixed values of the mixing-length parameter ($\alpha=1.8$) and the initial
hydrogen abundance ($X_{\mathrm{i}}=0.7$).  The grid covered masses in the range 0.7 to $2.4\,\mathrm{M}_\odot$ with a resolution
of $0.01\,\mathrm{M}_\odot$ and metallicities in the range $0.011\leq Z\leq 0.028$ with a resolution in $\log (Z/X)$ of 0.2 dex.
Figure \ref{fig1} shows the H-R diagram for models of near-solar metallicity ($Z_0=0.017$) for masses from 0.7 to
$2.0\,\mathrm{M}_\odot$. Models that are hotter than the approximate cool edge of the classical instability strip \citep{Saio98}
are colored gray; they are not expected to show solar-like oscillations because they do not have a significant convective
envelope. Many of these stars will show classical pulsations as $\delta$ Scuti, $\gamma$ Dor or roAp stars. However, we do
note that \citet{Antoci11} announced evidence for solar-like oscillations in a $\delta$ Scuti star.

The model parameters chosen, and the physics included in the models, do have an impact on the model frequencies obtained.
Several studies have already investigated the impact of the choice of these parameters, including composition, mixing, 
overshooting and diffusion \citep[e.g.][]{Monteiro02,Mazumdar05,Gai09}, so we have not investigated the breadth of this 
parameter space with our model grid. Our aim is to investigate the bulk behavior of asteroseismic observables across a 
wide range of evolutionary states, from the ZAMS to the tip of the RGB.

The adiabatic frequencies of every model were calculated using \textsc{ADIPLS} \citep{C-D08b}, adjusted to enable
proper sampling of the extremely high order eigenmodes that occur in red giants. Oscillation frequencies determined from
stellar models include more modes than can be observed. It is therefore necessary to determine asteroseismic parameters
from them with care so that they are directly comparable to the parameters measured from data. Here we outline our
approach. 

\subsection{Measuring $\Delta\nu$ and $\epsilon$}\label{MeasureLargeSepEps}

\begin{figure}
\epsscale{1.2}
\plotone{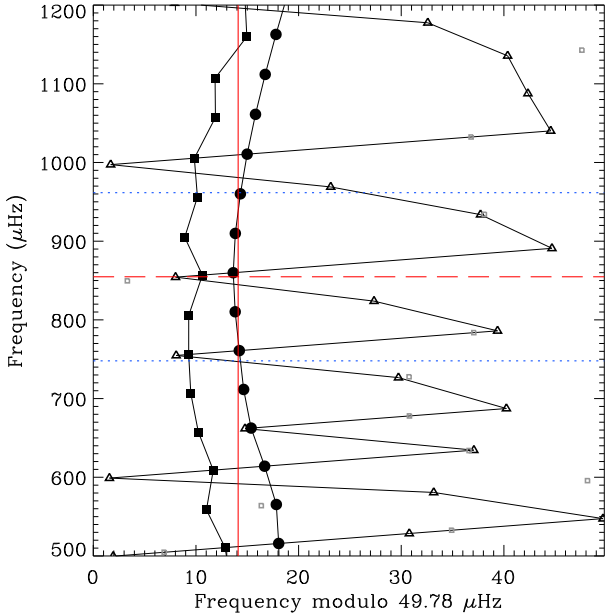}
\caption{An example of fitting the $l=0$ ridge of a subgiant model. This model has $M=1.0\,\mathrm{M}_{\odot}$, $Z_0=0.017$,
$\tau=10.78\,$Gyr. The $l=0$ modes are plotted with circles, $l=1$ with triangles and $l=2$ with squares. The vertical red line
shows the fit to the radial modes. The horizontal dashed red line indicates $\nu_{\mathrm{max}}$, and the dotted blue
lines indicate the FWHM of the Gaussian envelope used in the fit. Clearly visible is the distortion to the $l=1$ ridge
due to avoided crossings that would complicate the measurement of $\delta\nu_{01}$. Filtering the $l=2$ modes by
their mode inertia successfully removes the modes most affected by avoided crossings (open gray squares). }\label{fig2}
\end{figure}

\begin{figure}
\epsscale{1.2}
\plotone{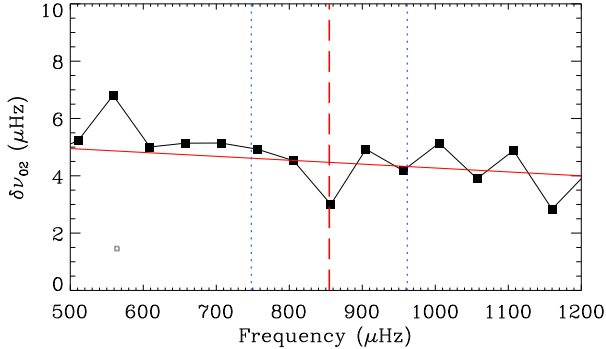}
\caption{An example of fitting the $l=2$ ridge according to equation~(\ref{smallSep}) for the same model as used in Figure
\ref{fig2}. Pairwise separations included in the fit are indicated by filled squares. Once again, filtering the $l=2$
modes by their mode inertia successfully removes from the fit the modes most affected by avoided crossings. The near-horizontal red line shows the fit to the pairwise separations. The dashed red line indicates $\nu_{\mathrm{max}}$, and
the dotted blue lines indicate the FWHM of the Gaussian envelope used in the fit.}\label{fig3}
\end{figure}

It is important that we treat the frequencies from models and data the same, as much as possible, so that the
comparisons between the two may be validly made. We therefore consider the observed characteristics of oscillations when
deciding on a method for fitting to frequencies that can be consistently applied to both models and observations.

The amplitudes of solar-like oscillations are modulated by an envelope that is approximately Gaussian. The peak of this
envelope is at $\nu_\mathrm{max}$, the frequency of maximum power. Since it is near $\nu_\mathrm{max}$ that the oscillations
have the most power, and are therefore the most easily observed, we chose to measure the oscillation properties about 
this point. To determine $\nu_\mathrm{max}$ for models we used the scaling relation \citep{Brown91,Kjeldsen95},
\begin{equation}
\frac{\nu_\mathrm{max}}{\nu_\mathrm{max,\odot}}=\frac{M/\mathrm{M_\odot}(T_\mathrm{eff}/\mathrm{T_{eff,\odot}})^{3.5}}{L/\mathrm{L_\odot}}.
\label{numaxScaling}
\end{equation}

We must then choose which model frequencies to include in our calculation of asteroseismic parameters. We could simply
take the frequencies within a specified range around $\nu_\mathrm{max}$, and fit to these frequencies. However,
this causes difficulties when, as $\nu_\mathrm{max}$ varies, frequencies at the top and bottom fall in and out of this
range from one model to the next. This will result in jumps in the derived quantities that are not physical. To overcome
this, we instead performed a weighted fit, using weights that decrease towards zero away from $\nu_\mathrm{max}$.
Inspired by the approximately Gaussian envelope of the oscillation amplitudes, we weighted the frequencies by a Gaussian
window centered on $\nu_\mathrm{max}$. The width of this Gaussian needs to be selected appropriately. 
The window should not be so wide as to include model frequencies that are unlikely to be observed. On the other hand,
a narrower window is sensitive to departures from the asymptotic relation as a result of acoustic glitches.
We have found that a full-width-at-half-maximum of $0.25\,\nu_\mathrm{max}$ is a good compromise.

To measure $\Delta\nu$ and $\epsilon$, we performed a weighted least-squares fit to the radial ($l=0$) frequencies as a
function of $n$. By equation~(\ref{asymp}), the gradient of this fit is $\Delta\nu$ and the intercept is
$\epsilon\Delta\nu$. An example of a fit to the $l=0$ frequencies of a stellar model is shown in Figure \ref{fig2} in 
\'echelle format, in which the frequencies are plotted against frequency modulo $\Delta\nu$. In the \'echelle diagram, 
frequencies that are separated by precisely $\Delta\nu$ will align vertically. The curvature in the $l=0$ and $l=2$ ridges 
indicates variation in either $\Delta\nu$ or $\epsilon$ (or both) as a function of frequency. Our choice of the fitting
method could potentially impact on our measured values of $\Delta\nu$ and $\epsilon$. We have tried different widths
for the Gaussian envelope and found that the measured value of $\Delta\nu$ does not vary significantly. However, there 
is a substantial change in the value of $\epsilon$ due to the influence of acoustic glitches \citep{Gough90,Houdek07}.
As the Gaussian is made wider, more orders contribute to the fit, averaging over the curvature in the \'echelle diagram.
Narrower Gaussians are more susceptible to curvature, which leads to a significant change in $\epsilon$, which in
extreme cases can approach a shift of 0.2 in $\epsilon$. This is a greater concern for higher-mass stars, for which
models exhibit greater curvature. Our chosen Gaussian window ($0.25\,\nu_\mathrm{max}$) is wide enough to average over
much of the curvature, while ensuring that no more frequencies are included in the fit to the models than can
reasonably be expected to be observed. More sophisticated approaches, involving a fit to the curvature, are beyond the 
scope of this paper.

\subsection{Measuring $\delta\nu_{02}$}
There are several small separations that can be measured. Only modes with $l\le3$ have been detected from solar-like
oscillations in stars other than the Sun, and in most cases, $l=3$ modes are not easily observed. The $l=1$
modes in subgiants can be significantly shifted in frequency due to avoided crossings, as exhibited by the model in 
Figure \ref{fig2}. Although the $l=2$ modes also
undergo avoided crossings, those modes that are bumped significantly in frequency have much greater inertia than
non-bumped modes and hence much smaller amplitudes. This contrasts with mixed $l=1$ modes, for which even strongly bumped
modes retain observable amplitudes. It follows that the $l=2$ ridge in the \'echelle diagram is quite well-defined, even
in subgiant stars. For this reason, it is the small separation $\delta\nu_{02}$, between modes of $l=0$ and 2, that we
have chosen to measure. To minimize the effect of bumped modes on our measurement of $\delta\nu_{02}$ from the models, we
ignored those $l=2$ modes (gray squares in Figure \ref{fig2}) whose inertia was significantly larger 
than that of adjacent non-bumped modes, typically by at least an order of magnitude.

Another consideration is that $\delta\nu_{02}$ decreases with frequency. This can be seen in the gradual decrease of the
separations between $l=0$ and $l=2$ modes, as shown in Figure \ref{fig3}. Guided by the Sun, for which $\delta\nu_{02}$
decreases approximately linearly with frequency \citep{Elsworth90}, we fitted the changing small separation by 
\begin{equation}
\delta\nu_{02} = \langle\delta\nu_{02}\rangle + k_{02}(\nu_{n,2}-\nu_{\rm{max}}),\label{smallSep}
\end{equation}
where $k_{02}$ is a slope to be determined by the fit. We weighted this fit by the same Gaussian envelope used in Section
\ref{MeasureLargeSepEps}. As for $\Delta\nu$, we did not find any significant change in the determined value of 
$\delta\nu_{02}$ when we varied the envelope width.

\section{Scaling relation for $\Delta\nu$}\label{scale}

\begin{figure}
\epsscale{1.2}
\plotone{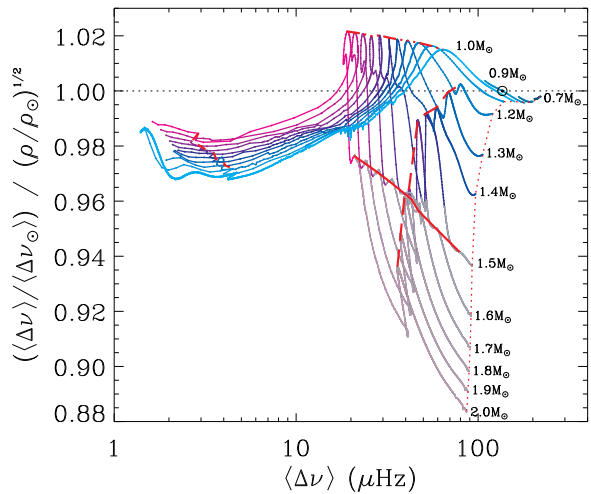}
\caption{The ratio $\Delta\nu/\Delta\nu_\odot$ to $(\rho/\rho_\odot)^{1/2}$ as a function of $\Delta\nu$ in models of
near-solar metallicity ($Z_0=0.017$) and mass range from 0.7 (green) to $2.0\,\mathrm{M}_\odot$ (magenta). Models that have
effective temperatures hotter than the approximate cool edge of the classical instability strip are shown in gray. The
zero age main sequence is indicated by the dotted red line. Sharp features in the tracks are indicated by the solid,
long-dashed, dot-dashed and short-dashed red lines, which, for reference, are also labeled as such in Figure
\ref{fig1}. The location of a solar model is marked by the Sun's usual symbol.}\label{fig4}
\end{figure}

\begin{figure}
\epsscale{1.2}
\plotone{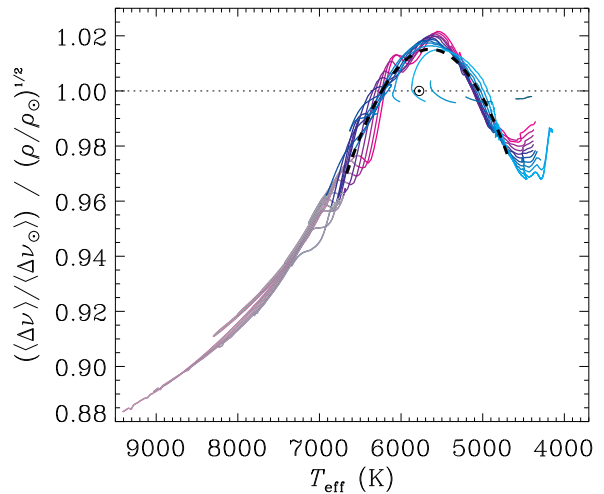}
\caption{The ratio $\Delta\nu/\Delta\nu_\odot$ to $(\rho/\rho_\odot)^{1/2}$ as a function of effective temperature,
$T_\mathrm{eff}$ in models of near-solar metallicity ($Z_0=0.017$) and mass range from 0.7 (green) to $2.0\,\mathrm{M}_\odot$
(magenta). The location of a solar model is marked by the Sun's usual symbol. The dashed black line shows the function given by equation \ref{scaling3}.}\label{fig5}
\end{figure}

\begin{figure}
\epsscale{1.2}
\plotone{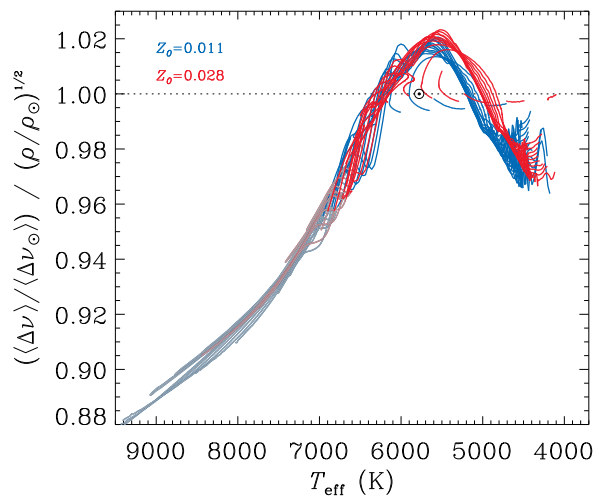}
\caption{The ratio $\Delta\nu/\Delta\nu_\odot$ to $(\rho/\rho_\odot)^{1/2}$ as a function of effective temperature,
$T_\mathrm{eff}$ in metal-poor ($Z_0=0.011$, blue) and metal-rich ($Z_0=0.028$, red) models and mass range from 0.7 to
$2.0\,\mathrm{M}_\odot$. The location of a solar model is marked by the Sun's usual symbol.}\label{fig6}
\end{figure}

\begin{deluxetable*}{lrrrrl}
\tabletypesize{\scriptsize}
\tablewidth{420pt}
\tablecaption{Measurements of $\Delta\nu$, $\delta\nu_{02}$ and $\epsilon$ from published frequency lists.\label{tbl-1}}
\tablehead{
\colhead{Star} &\colhead{$\nu_\mathrm{max}$} &\colhead{$\Delta\nu$} &\colhead{$\delta\nu_{02}$} & \colhead{$\epsilon$} & \colhead{Source}\\
& \colhead{($\mu$Hz)} & \colhead{($\mu$Hz)} & \colhead{($\mu$Hz)} & & }
\startdata
$\tau$\,Cet & 4500 & 169.47 $\pm$ 0.35 & 9.9 $\pm$ 2.0 & 1.45 $\pm$ 0.05 & \citet{Teixeira09} \\
$\alpha$\,Cen\,B & 4100 & 161.70 $\pm$ 0.24 & 10.8 $\pm$ 1.4 & 1.43 $\pm$ 0.04 & \citet{Kjeldsen05} \\
Sun & 3100 & 135.00 $\pm$ 0.11 & 8.88 $\pm$ 0.01 & 1.48 $\pm$ 0.02 & \citet{Broomhall09} \\
KIC\,6603624 (`Saxo') & \colhead{---} & 110.2 $\pm$ 0.6\tablenotemark{a} & 4.7 $\pm$ 0.2\tablenotemark{a} & \colhead{---} & \citet{Chaplin10} \\
$\alpha$\,Cen\,A & 2400 & 105.72 $\pm$ 0.27 & 6.44 $\pm$ 0.50 & 1.36 $\pm$ 0.05 & \citet{Bedding04} \\
HD\,52265 & 2090 & 98.22 $\pm$ 0.14 & 8.06 $\pm$ 0.32 & 1.33 $\pm$ 0.03 & \citet{Ballot11}\\
KIC\,3656476 (`Java') & \colhead{---} & 94.1 $\pm$ 0.6\tablenotemark{a} & 4.4 $\pm$ 0.2\tablenotemark{a} & \colhead{---} & \citet{Chaplin10} \\
$\mu$\,Ara & 2000 & 89.68 $\pm$ 0.19 & 5.56 $\pm$ 0.76 & 1.42 $\pm$ 0.04 & \citet{Bouchy05} \\
HD\,49933 (Scenario A) & 1760 & 85.54 $\pm$ 0.20 & 1.8 $\pm$ 1.1 & 1.54 $\pm$ 0.05 & \citet{Appourchaux08} \\
HD\,49933 (Scenario B) & 1760 & 85.53 $\pm$ 0.18 & 4.17 $\pm$ 0.53 & 1.06 $\pm$ 0.04 & \citet{Benomar09} \\
HD\,181420 (Scenario 1) & 1600 & 75.20 $\pm$ 0.32 & 6.34 $\pm$ 0.92 & 0.92 $\pm$ 0.09 & \citet{Barban09} \\
HD\,181420 (Scenario 2) & 1600 & 75.34 $\pm$ 0.25 & 1.26 $\pm$ 0.81 & 1.36 $\pm$ 0.07 & \citet{Barban09} \\
$\beta$\,Hyi & 1000 & 57.34 $\pm$ 0.20 & 5.32 $\pm$ 0.20 & 1.51 $\pm$ 0.06 & \citet{Bedding07} \\
KIC\,10920273 (`Scully') & 974 & 57.22 $\pm$ 0.13 & 4.86 $\pm$ 0.17 & 1.43 $\pm$ 0.04 & \citet{Campante11} \\
HD\,49385 & 1013 & 56.27 $\pm$ 0.19 & 4.20 $\pm$ 0.17 & 1.17 $\pm$ 0.06 & \citet{Deheuvels10} \\
Procyon (Scenario A) & 1000 & 55.55 $\pm$ 0.26 & 5.27 $\pm$ 0.52 & 0.76 $\pm$ 0.08 & \citet{Bedding10} \\
Procyon (Scenario B) & 1000 & 55.93 $\pm$ 0.22 & 2.98 $\pm$ 0.50 & 1.16 $\pm$ 0.07 & \citet{Bedding10} \\
KIC\,11026764 (`Gemma') & 900 & 50.49 $\pm$ 0.11 & 4.46 $\pm$ 0.29 & 1.30 $\pm$ 0.03 & \citet{Metcalfe10} \\
KIC\,10273246 (`Mulder') & 842 & 48.87 $\pm$ 0.14 & 4.14 $\pm$ 0.46 & 1.08 $\pm$ 0.05 & \citet{Campante11} \\
KIC\,11395018 (`Boogie') & 830 & 47.66 $\pm$ 0.14 & 4.53 $\pm$ 0.20 & 1.37 $\pm$ 0.05 & \citet{Mathur11} \\
KIC\,11234888 (`Tigger') & 675 & 41.88 $\pm$ 0.10 & 2.72 $\pm$ 0.31 & 0.99 $\pm$ 0.04 & \citet{Mathur11} \\
$\eta$\,Boo & 750 & 40.52 $\pm$ 0.09 & 3.52 $\pm$ 0.36 & 1.07 $\pm$ 0.04 & \citet{Kjeldsen03} \\
KIC\,4351319 (`Pooh') & 386 & 24.53 $\pm$ 0.07 & 2.32 $\pm$ 0.29 & 1.41 $\pm$ 0.04 & \citet{DiMauro11} \\
HD\,186355 (KIC\,11618103) & 106 & 9.32 $\pm$ 0.05 & 1.20 $\pm$ 0.05 & 1.25 $\pm$ 0.05 & \citet{Jiang11} \\
HR\,7349 & 28.2 & 3.44 $\pm$ 0.02 & 0.68 $\pm$ 0.04 & 1.05 $\pm$ 0.04 & \citet{Carrier10} \\
470 \em{Kepler }\em Giants & \colhead{---} & 2 -- 20\tablenotemark{a} & 0.2 -- 2.5\tablenotemark{a} & 0.6 -- 1.5\tablenotemark{a} & \citet{Huber10}
\enddata
\tablenotetext{a}{Value taken directly from source without fitting to a frequency list.}
\end{deluxetable*}

Before discussing the C-D~diagram, we shall first explore the scaling relation for the large separation. As previously mentioned, it is well-established that $\Delta\nu$ is approximately proportional to the square root
of the mean density of the star. This leads to the scaling relation
\begin{equation}
\rho \approx \left({\Delta\nu \over \Delta\nu_\odot}\right)^2 \rho_\odot,  \label{scaling}
\end{equation}
where $\rho$ is the mean density of a star, $\Delta\nu$ is its large separation, and $\rho_\odot$ and $\Delta\nu_\odot$
are the corresponding quantities for the Sun. This rather simple relation allows an estimate of the mean
density of any star for which $\Delta\nu$ can be measured. 

Given its widespread use, it is important to validate equation~(\ref{scaling}). Observationally, this can only be done 
for the handful of stars whose masses and radii have been measured independently. However, we can at least ask how 
accurately equation~(\ref{scaling}) is followed by models. Detailed model calculations have confirmed the
scaling relation for main-sequence stars of low mass \citep[$<1.2\,\mathrm{M}_\odot$;][]{Ulrich86}. \citet{Stello09} investigated
the relation up to $2.0\,\mathrm{M}_\odot$ by calculating $\Delta\nu$ from the integral of the sound speed,
$[2\int\mathrm{d}r/c]^{-1}$. However, they only showed the relation between $[2\int\mathrm{d}r/c]^{-1}$ and $\Delta\nu$
measured from the frequencies for a restricted set of models. \citet{Basu10} have also compared $\Delta\nu$ as
calculated from the scaling relation to the average $\Delta\nu$ from calculated model frequencies for a representative
subset of models of various masses and evolutionary states. While they found the scaling relation broadly applies, they
did not investigate smaller scale deviations. With our calculations of $\Delta\nu$, we are able to investigate the
validity of the scaling relation more directly over a broader range of masses and evolutionary states.

\begin{figure*}
\epsscale{1.2}
\plotone{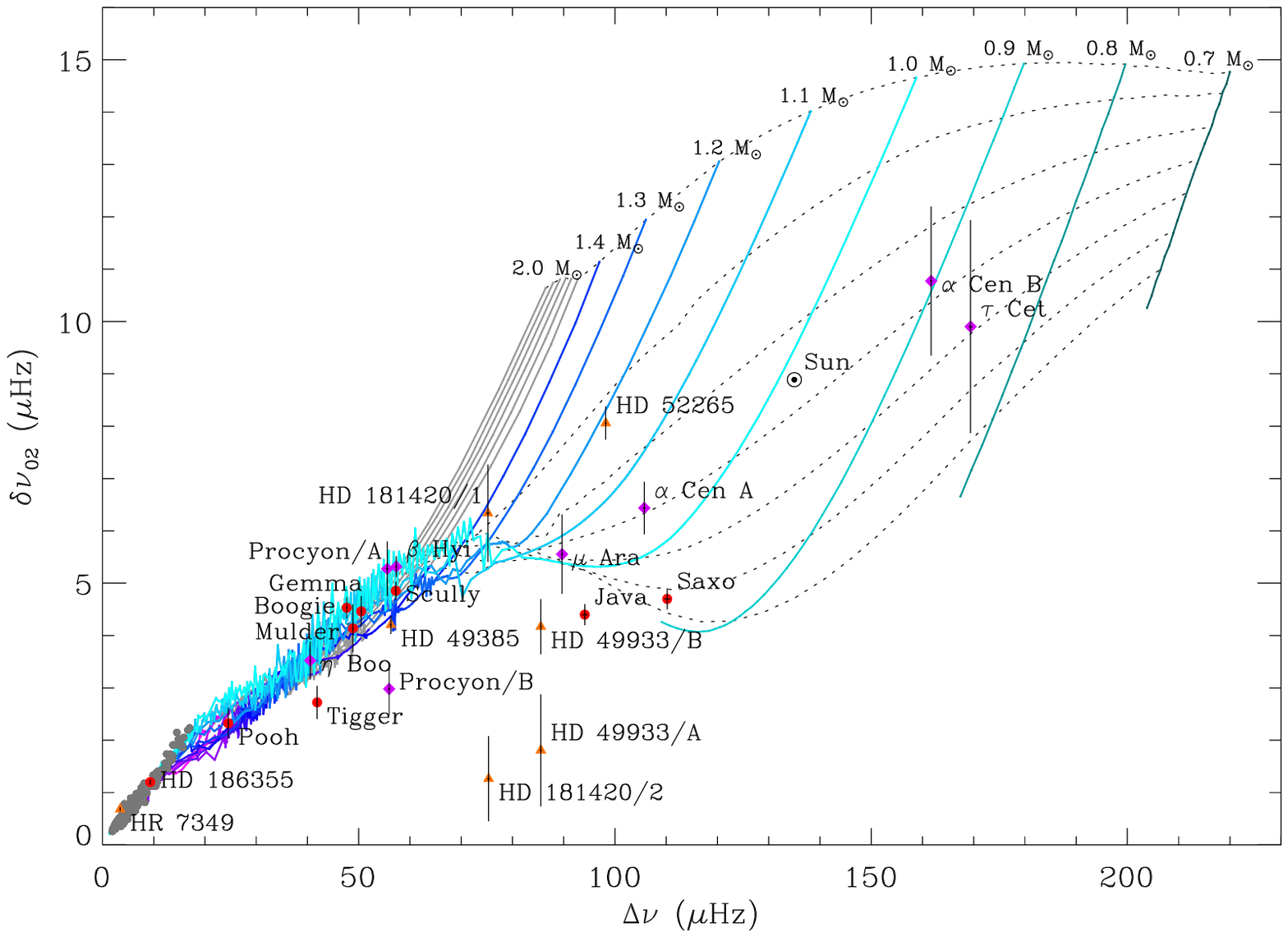}
\caption{C-D~diagram, with model tracks for near-solar metallicity ($Z_0=0.017$). Tracks increase in mass by
$0.1\,\mathrm{M}_\odot$ from $0.7\,\mathrm{M}_\odot$ to $2.0\,\mathrm{M}_\odot$ as labeled. The section of the evolutionary tracks in
which the models have a higher $T_\mathrm{eff}$ than the approximate cool edge of the classical instability strip
\citep{Saio98} are gray; they are not expected to show solar-like oscillations. Dashed black lines are isochrones,
increasing by 2\,Gyr from 0\,Gyr (ZAMS) at the top to 12\,Gyr at the bottom. Stars shown, as labeled, were observed by
either {\it CoRoT} (orange triangles), {\it Kepler} (red circles) or from the ground (purple diamonds). Gray circles are {\it
Kepler} red giants \citep{Huber10}. The Sun is marked by its usual symbol.}\label{fig7}
\end{figure*}

We show in Figure \ref{fig4} the ratio of $\Delta\nu/\Delta\nu_\odot$ to $(\rho/\rho_\odot)^{1/2}$ as a function of
$\Delta\nu$ for models of mass 0.7--2.0~$\mathrm{M}_\odot$ and initial metallicity $Z_0=0.017$, close to the solar value. 
The section of the evolutionary tracks for which the models have a higher effective temperature than the approximate cool
edge of the classical instability strip \citep{Saio98} are colored gray. For reference, four `features' in the evolutionary
tracks of Figure \ref{fig4} are indicated in the H-R diagram in Figure \ref{fig1}. These features relate to the
onset of a convective envelope (solid line) and of a convective core (long dashes), the Hertzsprung gap towards the base of
the RGB (dot-dashes) and the so-called `bump'  (short dashes) where the hydrogen-burning shell burns through the
discontinuity in molecular weight left from the deep convective envelope (`first dredge up') during the early ascent of
the RGB \citep{Thomas67, Iben68}.

Since we are testing the scaling relation in models, it is appropriate that we use the value of $\Delta\nu_\odot$ from a solar
model rather than the observed value in the Sun. These values differ due to the offset between observed and computed
oscillation frequencies. This offset is known to arise from an improper modeling of near-surface layers \citep{C-D88b, Dziembowski88, C-D96,
C-D97} and is presumably a problem for other stars as well \citep{Kjeldsen08}. The offset increases with frequency, at
least in the Sun, and so affects the large separation, with $\Delta\nu$ being $\sim1$\% greater in solar models than
observed \citep{Kjeldsen08}. We therefore adopted $\Delta\nu_\odot=135.99\,\mu\mathrm{Hz}$, derived from a fit to frequencies of the 
well-studied model~S of \citet{C-D96}. We note that the surface offset also has a significant effect on $\delta
\nu_{02}$ \citep{Roxburgh03}.

From Figure \ref{fig4} we can see that the scaling relation holds quite well for lower-mass main-sequence
stars, but there is a significant deviation at other masses and evolutionary states. These deviations can be as large as 3\%
for low-mass red giants and are over 10\% within the instability strip. 

In Figure \ref{fig5} we show the ratio against effective temperature, $T_{\mathrm{eff}}$. From this we see
that the deviation from the scaling relation is predominantly a function of effective temperature, which
suggests that the scaling relation could be improved by incorporating a function of $T_{\mathrm{eff}}$. A rough
approximation to the models suggests a variation of the scaling relation of the form
\begin{equation}
{\rho \over \rho_\odot} = \left({\Delta\nu \over \Delta\nu_\odot}\right)^2 (f(T_\mathrm{eff}))^{-2}, \label{scaling2}
\end{equation}
where
\begin{equation}
f(T_\mathrm{eff}) =
-4.29\left({T_\mathrm{eff}\over10^4\,\mathrm{K}}\right)^2+4.84\left({T_\mathrm{eff}\over10^4\,\mathrm{K}}\right)-0.35 \label{scaling3}
\end{equation}
for stars with temperatures between 4700 and 6700\,K, except main-sequence stars below $\sim1.2\,\mathrm{M}_\odot$,
for which it seems best to set $f(T_\mathrm{eff})=1$. The function in equation~(\ref{scaling3}) is the black dashed line 
shown in Figure \ref{fig5}. Making this adjustment improves the accuracy of the scaling relation to approximately 1\% 
for these models. 

Does metallicity also have an impact on the scaling relation? In Figure \ref{fig6} we show the ratio of
$\Delta\nu/\Delta\nu_\odot$ to $(\rho/\rho_\odot)^{1/2}$ as a function of temperature for metal-poor ($Z_0=0.011$,
$-0.2$ dex) and metal-rich ($Z_0=0.028$, $+0.2$ dex) models. We see that metallicity has little effect except for red
giants, for which there is a slight dependence.

After all this, however, we must keep in mind the surface correction. As discussed above, there is a frequency-dependent
offset between models and observations that affects $\Delta\nu$. A full investigation of this must await improvements to
the models, but meanwhile we recommend that using equation~(\ref{scaling2}) (or equation~(\ref{scaling})) to estimate stellar
densities from observed frequencies be based on the {\em observed} value of $\Delta\nu_\odot$ ($135.0\,\mu\mathrm{Hz}$).

\section{The C-D Diagram}\label{CD}

\begin{figure*}
\epsscale{1.2}
\plotone{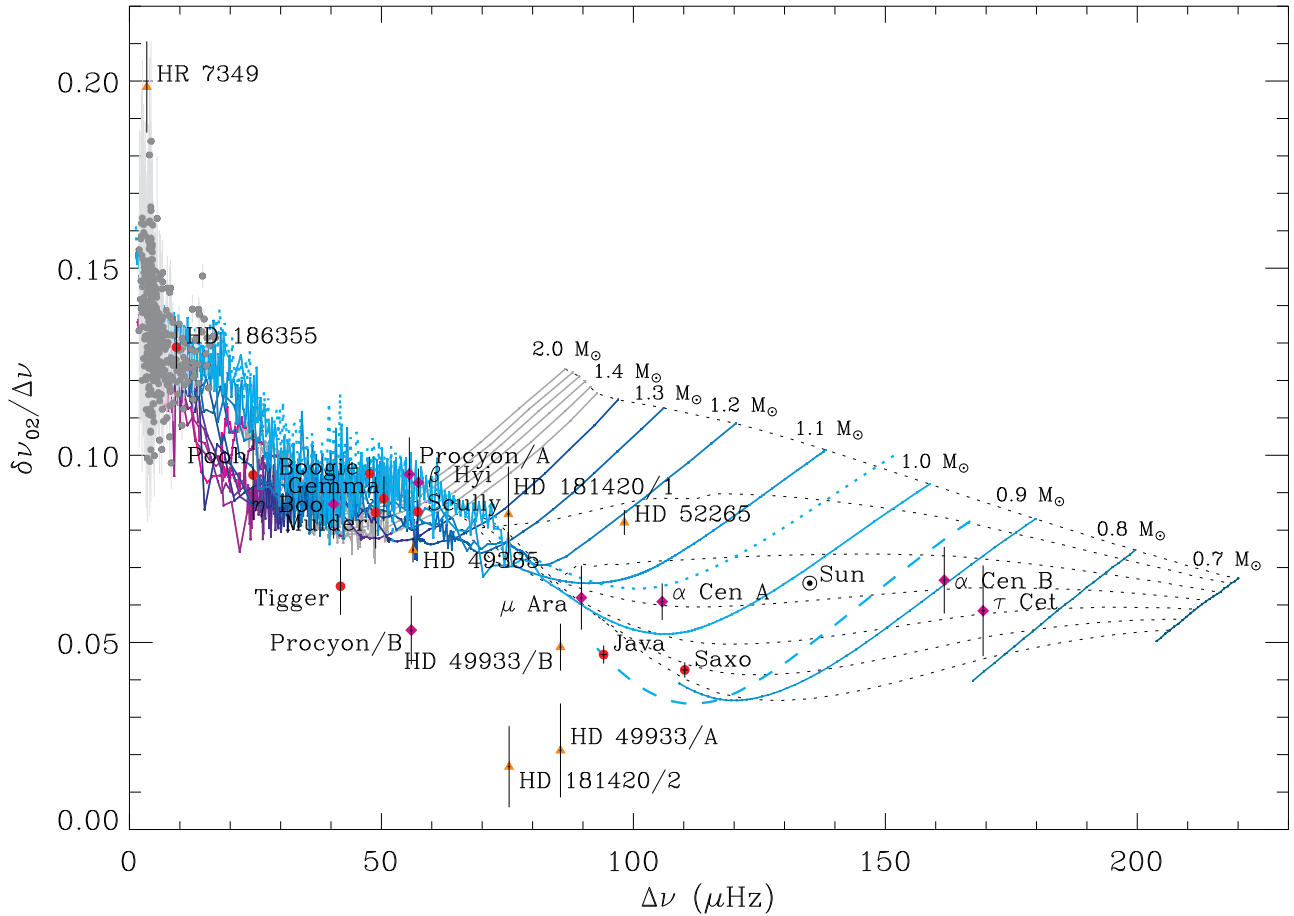}
\caption{Modified C-D~diagram using the ratio $\delta\nu_{02}$/$\Delta\nu$, with near-solar metallicity ($Z_0=0.017$)
model tracks. Also shown are tracks for metal-poor ($Z_0=0.011$; dotted) and metal-rich ($Z_0=0.028$; dashed) solar-mass
models. Other colors, lines and symbols are the same as for Figure \ref{fig7}.}\label{fig8}
\end{figure*}

\begin{figure*}
\epsscale{1.2}
\plotone{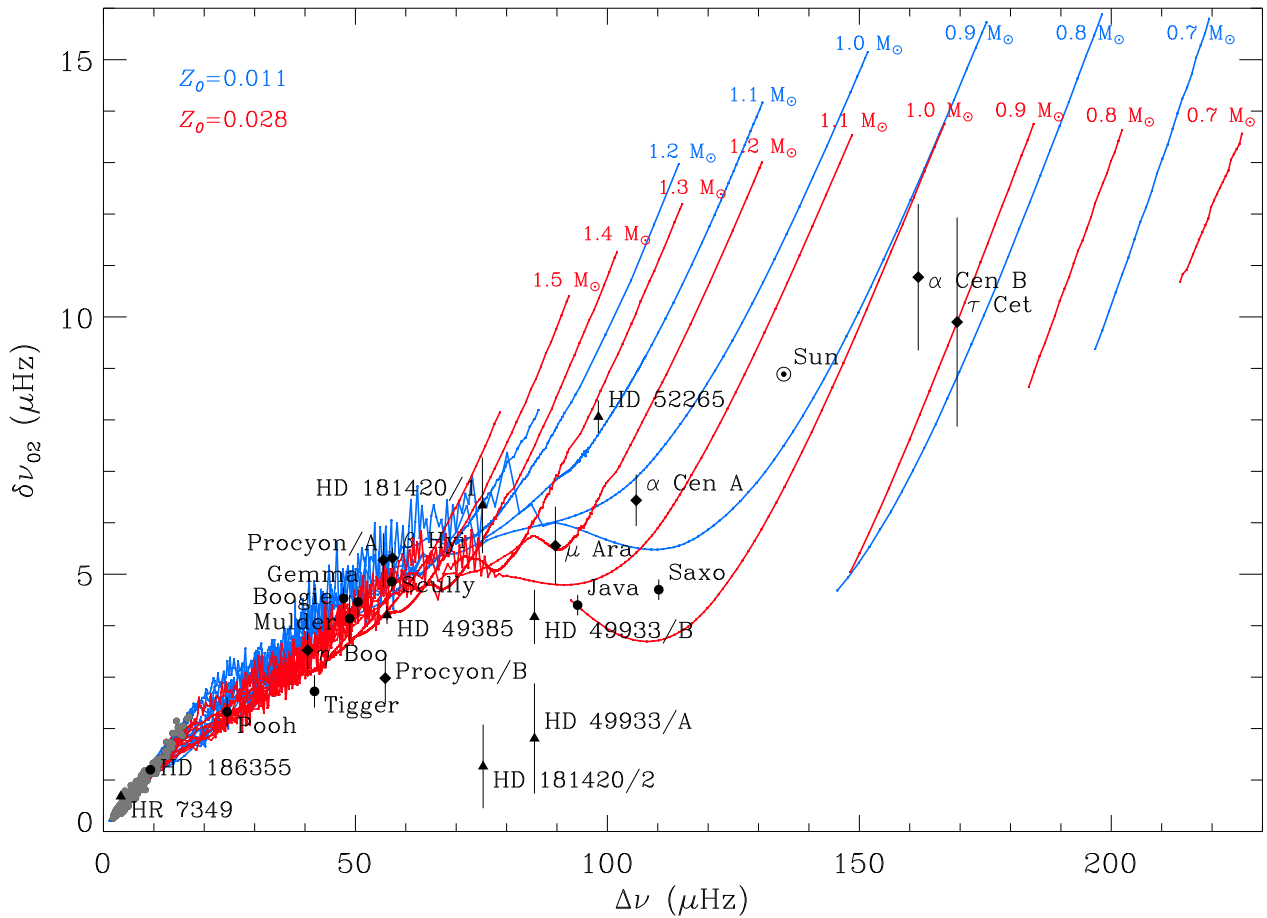}
\caption{C-D~diagram, with model tracks for stars that are metal poor ($Z_0=0.011$, blue) and metal rich
($Z_0=0.028$, red) equivalent to a range in [Fe/H] of 0.4 dex. For clarity isochrones and model tracks with a higher
$T_\mathrm{eff}$ than the approximate cool edge of the classical instability strip are not shown. Black symbols are for
stars observed by {\it CoRoT} (triangles), {\it Kepler} (circles) and from the ground (diamonds). Gray circles are {\it
Kepler} red giants \citep{Huber10}.The Sun is marked by its usual symbol. }\label{fig9}
\end{figure*}

The C-D~diagram is shown in Figure \ref{fig7}. The solid lines show the evolution of $\Delta\nu$ and $\delta\nu_{02}$ for models
with a metallicity close to solar and various masses. Stars evolve from the top-right of the diagram to the bottom-left.
Isochrones are also shown as dashed lines.
In applying this diagram it should be recalled that both $\Delta \nu$
and $\delta \nu_{02}$ are affected by the errors in the treatment of the
near-surface layers.
Modeling indicates that, e.g., the ratio of $\delta\nu_{02}$ to $\Delta\nu$ is
less sensitive to surface layer effects \citep{Roxburgh03,OtiFloranes05,Mazumdar05}.
Figure \ref{fig8} shows a modified C-D~diagram, which uses this frequency-separation ratio,
although the surface dependency remains in $\Delta\nu$. The isochrones are close to horizontal in this figure, showing 
that this ratio is an effective indicator of age.
We note that $\Delta\nu$ is typically measured from the $l=1$ modes when calculating this ratio, but since the 
$l=1$ modes depart significantly from the asymptotic relation for more evolved stars, we have determined 
$\Delta\nu$ using only $l=0$ modes. In the absence of avoided crossings, the difference between $\Delta\nu$ as 
measured from $l=0$ and $l=1$ modes is small, so we expect this change in the definition of the
ratio $\delta\nu_{02}$/$\Delta\nu$ to have little impact.

Variations on the C-D~diagram may be constructed by using different small separations in place of $\delta\nu_{02}$.
\citet{Mazumdar05} and \citet{Montalban10} have investigated the C-D~diagram using $\delta\nu_{01}$ for main-sequence 
and RGB stars, respectively. For subgiants, $\delta\nu_{01}$ becomes poorly defined due to avoided crossings causing 
a major departure from the asymptotic relation 
\citep[see Figure \ref{fig2} and][for examples]{Metcalfe10,Campante11,Mathur11}. We have therefore not 
considered the $\delta\nu_{01}$--$\Delta\nu$ C-D~diagram here.

The C-D~diagram is clearly most useful for main-sequence stars, particularly for masses $<\!\!1.5\,\rm{M}_\odot$, for
which the evolutionary tracks are well separated. As stars evolve off the main sequence, their tracks converge for the
subgiant and red-giant evolutionary stages. This convergence of the tracks means that the C-D~diagram is not a good
discriminant of age and mass for these stars. This behavior of the model tracks is consistent with early results of red
giants observed by {\it Kepler} \citep{Bedding10c,Huber10} and the modeling results of \citet{Montalban10}, for which it was found that $\delta\nu_{02}$ is an almost
fixed fraction of $\Delta\nu$. This also explains the observation by \citet{Metcalfe10}, when modeling the {\it Kepler}
subgiant KIC\,11026764, that including $\delta\nu_{02}$ in the fit to the models did not provide an additional
constraint beyond that provided by $\Delta\nu$.

To compare the models with observations we have measured $\Delta\nu$, $\delta\nu_{02}$ and $\epsilon$ from the published
frequency lists of 20 stars using the methods outlined in Section \ref{Measure}. The method described above was used for
calculating $\delta\nu_{02}$ except that, apart from the Sun, the data did not justify the inclusion of $k_{02}$ as an
extra parameter in the fit to the frequencies. We have therefore kept $k_{02}$ fixed at the solar value ($-0.0022$) when
fitting the other stars. Doing so did not significantly affect the measured value of $\delta\nu_{02}$, or its
uncertainty. The measured values are listed in Table \ref{tbl-1}, along with the values adopted for $\nu_\mathrm{max}$ when
fitting the frequencies of each star. Note that for two {\it Kepler} main-sequence stars and 470 red giants we used published $\Delta\nu$ and $\delta\nu_{02}$ values.

For three of the stars, the correct mode identification is ambiguous, and so we report both of the possible scenarios.
In HD\,49933, Scenario B is now accepted as the correct identification \citep{Benomar09}, and Scenario 1 in HD\,181420
seems most likely \citep{Barban09,Mosser09,Bedding10b}. With Procyon the situation is not yet settled, as we discuss
below.

In comparing the models with observations, it is clear that the positions of main-sequence
stars in the C-D~diagram can constrain their masses and ages. Given the typical uncertainties in the
observed values of $\Delta\nu$ and $\delta\nu_{02}$, the mass can be determined to a precision of a few percent, assuming that
other model parameters, such as the mixing-length parameter, are valid for the stars. Age is more difficult to
constrain, owing to the considerably larger fractional uncertainty in $\delta\nu_{02}$. Extreme cases are $\alpha$~Cen~B and
$\tau$~Ceti, which have large uncertainties in $\delta\nu_{02}$ due to having few reported $l=0,2$ frequency pairs. It is
worth noting that the published {\it Kepler} stars already have quite small uncertainties in the small separation, and
this will further improve as more data are collected.

Until now, we have only considered the C-D~diagram for models with near-solar metallicity. As shown by previous studies,
the metallicity has a major impact on the tracks, as does the physics in the models used \citep{Ulrich86, Gough87,
C-D88,Monteiro02,OtiFloranes05,Mazumdar05,Gai09}. The C-D~diagram for a series of metal-poor models ($Z_0=0.011$, blue)
and metal-rich models ($Z_0=0.028$, red) are plotted in Figure \ref{fig9}. For metal-poor models, the tracks for a given
mass shift up and to the left, and in the opposite direction for metal-rich models. The size of the shift underlines the
importance of obtaining spectroscopic abundances for asteroseismic targets.

For subgiants, where the tracks converge for different masses, the position of the star could largely depend on
metallicity. However, the presence of mixed $l=2$ modes can affect the measured small separation. Furthermore, the expected
shift in the small separation due to metallicity differences is similar in magnitude to the present measurement
uncertainties in the small separation, limiting the usefulness of the small separation as a proxy for metallicity in
subgiants.

\section{The $\epsilon$~Diagram}\label{Epsilon}

\begin{figure*}
\plotone{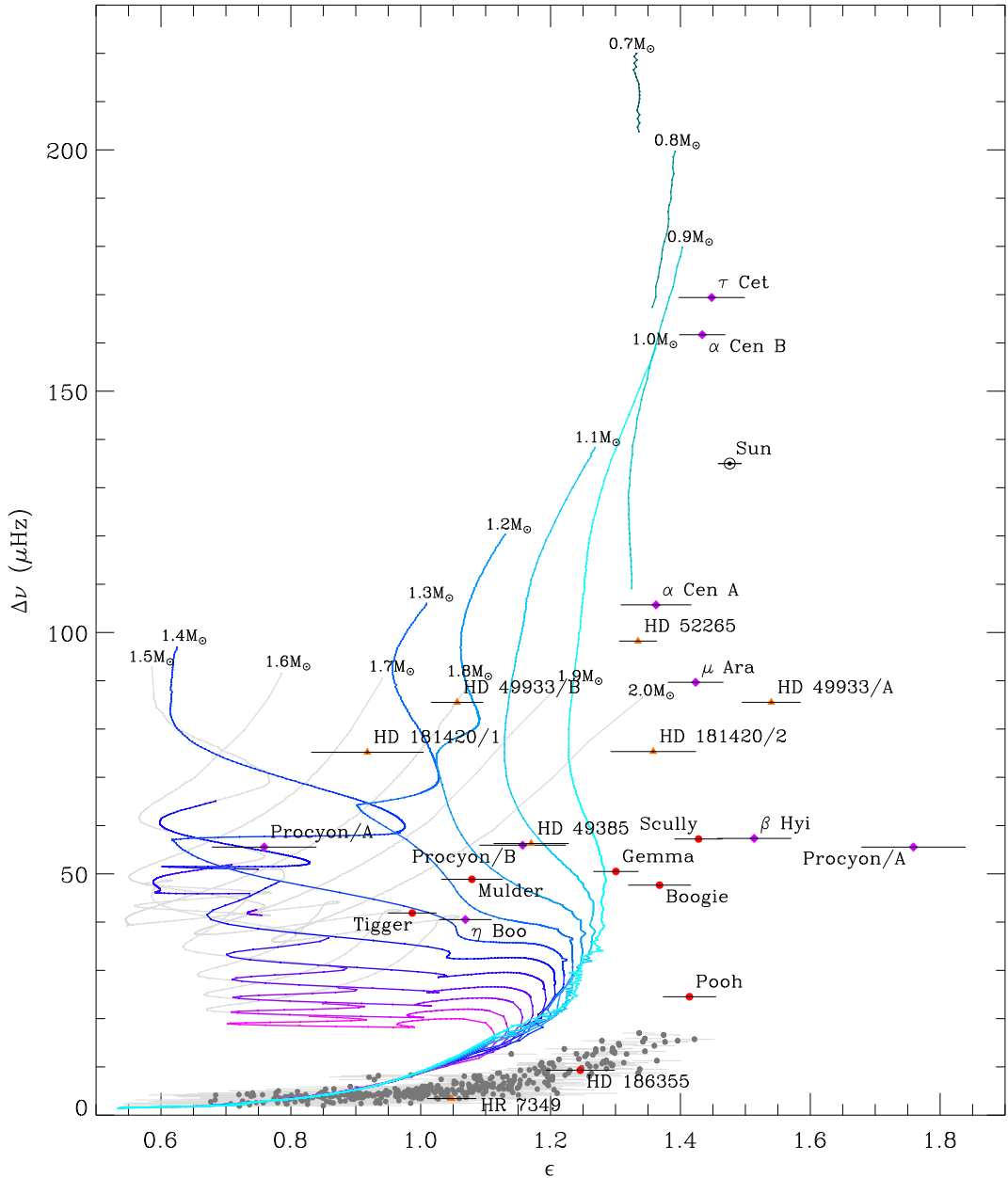}
\caption{The $\epsilon$~diagram, with near-solar metallicity ($Z_0=0.017$) model tracks. Tracks increase in mass by
$0.1\,\mathrm{M}_\odot$ from $0.7\,\mathrm{M}_\odot$ to $2.0\,\mathrm{M}_\odot$ (green to magenta lines). The section of the evolutionary tracks
hotter than the cool edge of the classical instability strip are gray. For clarity, isochrones are not shown. Symbols
for stars are the same as for Figure \ref{fig7}.}\label{fig10}
\end{figure*}

\begin{figure*}
\plotone{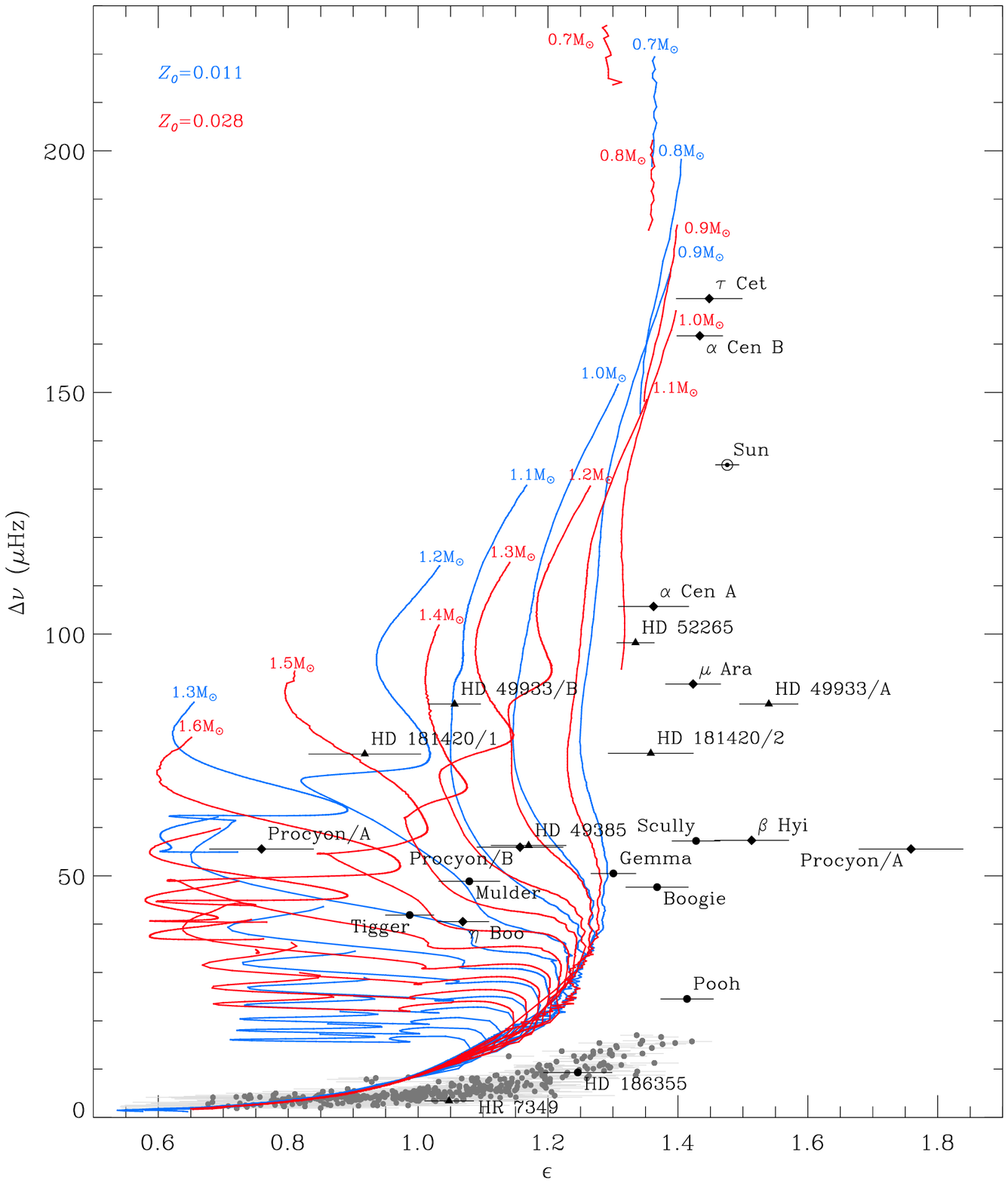}
\caption{The $\epsilon$~diagram, with model tracks for stars that are metal poor ($Z_0=0.011$, blue) and metal rich
($Z_0=0.028$, red). For clarity model tracks with a higher $T_\mathrm{eff}$ than the approximate cool edge of
the classical instability strip are not shown. Symbols for stars are the same as for Figure \ref{fig9}.}\label{fig11}
\end{figure*}

Having extended the C-D~diagram beyond the main sequence, we have found that the evolutionary tracks converge for stars
of different masses during the subgiant and red-giant phases. We now discuss the $\epsilon$~diagram, which breaks this 
degeneracy to some extent.

Figure \ref{fig10} shows the $\epsilon$~diagram with evolutionary tracks for models of mass 0.7--2.0 M$_\odot$ and
$Z_0=0.017$. Stars evolve from the top to the bottom in this diagram. Unlike in the C-D~diagram, the evolutionary tracks
in the $\epsilon$~diagram remain well separated for subgiants. This raises the possibility of using this diagram to
constrain mass and age. However, some difficulties arise that make this challenging, as we now discuss.

There can be a large uncertainty in the measurement of $\epsilon$, as is apparent for several stars shown in Figure
\ref{fig10}. This cannot be readily overcome by obtaining higher quality data because it is often due to the intrinsic
curvature of the $l=0$ ridge in the \'echelle diagram. To resolve this, it may be necessary to fit to this curvature,
although it may be difficult to do this consistently between models and observations in which only a few radial orders 
are observed.

The value of $\epsilon$ from observations may also be ambiguous by $\pm1$ since the radial order, $n$, of the modes is unknown
(unlike for models). For the stars considered here, it seems that only Scenario A of Procyon has an ambiguous value of
$\epsilon$.

An important feature of the $\epsilon$~diagram is the well-known offset between observed and computed oscillation
frequencies, as mentioned earlier, which manifests itself as an offset in $\epsilon$. This is the reason that the
observed values of $\epsilon$ in Figure \ref{fig10} are systematically offset from the models. One way to address
this issue is to correct the model frequencies empirically, as suggested by \citet{Kjeldsen08}. A more satisfactory
approach would naturally be to improve the modeling of the near-surface layers.

As in the C-D~diagram, metallicity has an impact on the evolutionary tracks in the $\epsilon$~diagram.
Diagrams for metal-poor stars ($Z_0=0.011$) and metal-rich stars ($Z_0=0.028$) are shown in Figure \ref{fig11}. The variation
in the position of  tracks due to metallicity further emphasizes the importance of supporting spectroscopic
measurements.

For red giants the evolutionary tracks are seen to converge and become independent of mass and metallicity. 
It has previously been suggested that the near-surface offset is negligible in red giants
\citep{Gilliland10,DiMauro11,Jiang11}. This was based upon finding model frequencies that were a close fit to
observed frequencies without any need for a correction. However, Figures \ref{fig10} and \ref{fig11} clearly show an
offset between the models and observations. We suggest that the published models referenced above do not agree with
observations as well as first thought. While the frequencies may appear to agree, the models may have a large separation
slightly greater than than the observations. Due to the low frequencies of red giant oscillations and the few orders
observed this effect is subtle, but if more orders were observed the discrepancy would become clear. We have
measured $\Delta\nu$ and $\epsilon$ consistently in models and observations and conclude that the near-surface offset is
significant for red giants.

Finally, we address whether $\epsilon$ really does discriminate between different stellar masses. Models indicate that
it does, but given the significant contribution to $\epsilon$ from near-surface layers, which are presently poorly
modeled, it is not yet known how the near-surface offset varies with mass, age or metallicity. It is therefore 
important to verify that stars of higher mass really do have lower values of $\epsilon$, as the models predict. To do 
this, we consider the relative positions of stars with known masses in the $\epsilon$~diagram.

\begin{deluxetable}{llrrr}
\tabletypesize{\scriptsize}
\tablewidth{260pt}
\tablecaption{Measured $\epsilon$, modelled masses and spectroscopic metallicities of subgiants \label{tbl-2}}

\tablehead{\colhead{KIC} & \colhead{Name} & \colhead{$\epsilon$} &\colhead{$M$} & \colhead{[Fe/H]}\\
& & & \colhead{($\mathrm{M}_\odot$)} & \colhead{(dex)}}
\startdata
    & $\beta$ Hyi & $1.51\pm0.06$ & $1.08\pm0.03$ & $-0.10\pm0.07$ \\
10920273 & Scully & $1.43\pm0.04$ & $1.13\pm0.05$ & $-0.03\pm0.08$ \\
11395018 & Boogie & $1.37\pm0.05$ & $1.36\pm0.07$ & $+0.27\pm0.09$ \\
11026764 & Gemma  & $1.30\pm0.03$ & 1.13 or 1.23  & $+0.02\pm0.06$ \\
10273246 & Mulder & $1.08\pm0.05$ & $1.43\pm0.04$ & $-0.10\pm0.07$ \\
     & $\eta$ Boo & $1.07\pm0.04$ & 1.64---1.75   & $+0.30\pm0.05$ \\
11234888 & Tigger & $0.99\pm0.04$ & $1.59\pm0.11$ & \colhead{---}
\enddata
\end{deluxetable}

Let us consider $\beta$ Hyi, $\eta$ Boo and the five {\it Kepler} subgiants, since it is for subgiants that $\epsilon$ 
is potentially most useful. Masses for $\beta$ Hyi and $\eta$ Boo have been estimated from models by \citet{Brandao11} 
and \citet{DiMauro03}, respectively. The metallicities adopted for this modeling were $-0.10\pm0.07$ for $\beta$ Hyi
\citep{Bruntt10} and $+0.30\pm0.05$ for $\eta$ Boo \citep{Taylor96}. KIC\,11026764 (Gemma), was modeled using the 
individual mode frequencies as constraints by \citet{Metcalfe10}. The four other {\it Kepler} subgiants were modeled by
\citet{Creevey11} using $\Delta\nu$ and $\nu_\mathrm{max}$ as seismic constraints. The masses determined from the 
modeling and the spectroscopic metallicities are given in Table \ref{tbl-2}, along with the the measured values of 
$\epsilon$. How do these masses and metallicities compare with the observed values of $\epsilon$?

In general, we do see a trend of decreasing $\epsilon$ with increasing mass, as expected from the evolutionary
tracks in Figure \ref{fig10}. The only deviations from this trend are KIC\,11395018 and $\eta$ Boo, both of which 
are substantially more metal-rich than the other subgiants. Increased metalicity results in a larger $\epsilon$ 
for a star of a given mass (Figure \ref{fig11}), so it is no surprise that KIC\,11395018 and $\eta$ Boo have a 
slightly larger $\epsilon$ than KIC\,11026764, despite being more massive.
This qualitative comparison shows that $\epsilon$ does depend on fundamental stellar parameters, such as 
mass and metallicity and confirms that $\epsilon$ is useful as an additional asteroseismic parameter.

\section{Using the diagrams for mode identification}\label{modeID}

\begin{figure}
\epsscale{1.2}
\plotone{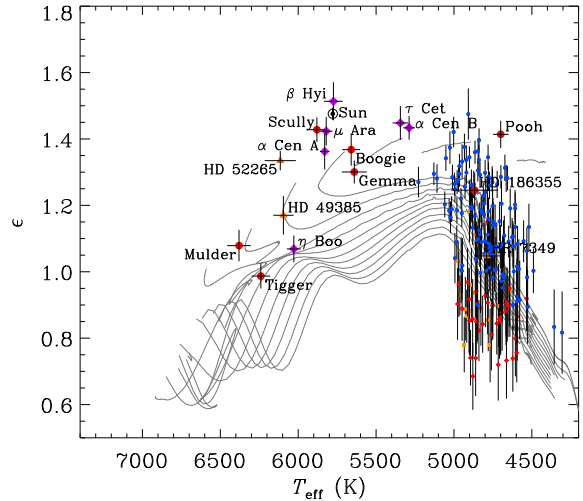}
\caption{$\epsilon$ as a function of effective temperature for models with $Z_0=0.017$ (gray lines) and observations of stars with secure
mode identifications. Symbols for main-sequence and subgiant stars are the same as for Figure \ref{fig7}. Red giants which
have been identified as hydrogen-shell burning RGB stars are indicated by blue circles, red clump stars by red
diamonds, and secondary clump stars by orange squares \citep{Bedding11}.}\label{fig12}
\end{figure}

\begin{figure}
\epsscale{1.2}
\plotone{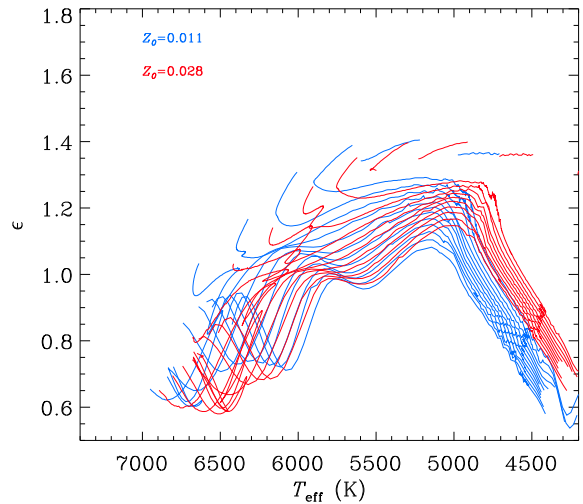}
\caption{$\epsilon$ as a function of effective temperature for models with $Z_0=0.011$ (blue) and $Z_0=0.028$ (red).}\label{fig13}
\end{figure}

\begin{figure}
\epsscale{1.2}
\plotone{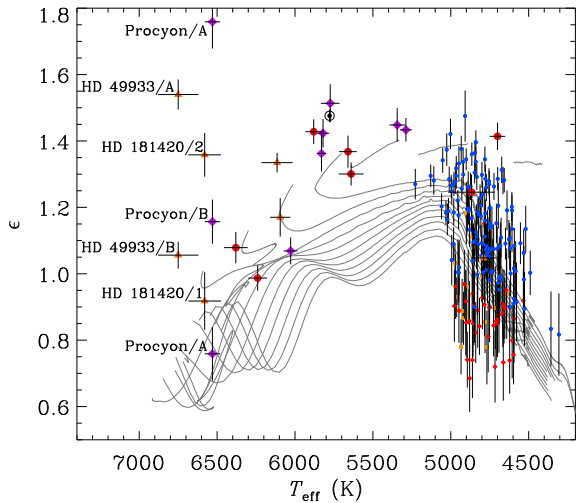}
\caption{Same as Figure \ref{fig12} with the addition of stars with ambiguous mode identifications.}\label{fig14}
\end{figure}

It has been difficult to establish the correct mode identification in F stars. Short mode lifetimes in these stars
result in large linewidths, blurring the distinction between the $l$=0,2 and $l$=1,3 ridges. It is common for both
possible mode scenarios to be fitted. Usually, it is noted that one of these identifications may be statistically more
likely, although the alternative cannot be ruled out. Indeed, the initially favored `Scenario A' of HD\,49933
\citep{Appourchaux08} was, with more data, found to be less likely than `Scenario B' \citep{Benomar09}.
\citet{Bedding10} favored the `Scenario B' identification of the F5 subgiant Procyon. We note, however, that
\citet{Huber11} favored `Scenario A' based upon their analysis of the combined ground-based radial velocity and MOST
space-based photometric observations. In modeling Procyon, \citet{Dogan10} found Scenario A to be least problematic from
a modeling perspective, but could not rule out Scenario B.

\citet{Bedding10b} suggested that $\epsilon$ may be able to resolve this problem. They reasoned that, if $\epsilon$
varies slowly with stellar parameters then by scaling the frequencies of one star whose identification is clear and
comparing to those of another should reveal the correct identification of the second. However, the $\epsilon$~diagrams 
presented in this paper cast doubt upon the validity of the assumption that $\epsilon$ varies slowly with stellar 
parameters. In some cases, $\epsilon$ is seen to vary quite rapidly as the star evolves, particularly for higher-mass 
subgiants such as Procyon. Nevertheless, we have found that $\epsilon$ is still a useful quantity for determining the 
most plausible identification, especially if we plot it against effective temperature.

Figure \ref{fig12} shows the relation between $\epsilon$ and $T_\mathrm{eff}$ for models, and for stars in which the mode
identification is secure. In this figure the model tracks for different masses have much less spread than when
plotted against $\Delta\nu$ (Figure \ref{fig10}). Once again, the near-surface offset is apparent, with the models having a systematically lower
$\epsilon$ than the observed stars. As shown in Figure \ref{fig13}, there is little metallicity dependence
in this relationship, except for red giants. Therefore, given the effective temperature of a star, $\epsilon$ can be
very useful in deciding the correct mode identification.

In Figure \ref{fig14} we again plot $\epsilon$ against $T_\mathrm{eff}$, but this time adding three F stars for
which mode identifications are ambiguous: Procyon, HD\,49933 and HD\,181420. For HD\,49933, as previously mentioned,
Scenario~B is now considered to be correct, and indeed this is the identification that falls along the observed
$\epsilon$--$T_\mathrm{eff}$ trend. Scenario~1 is the preferred identification in HD\,181420, and again this matches
the trend. Unfortunately the situation in Procyon is not completely clear. The trend with which we hope to
identify the correct $\epsilon$ is still loosely defined due to the scarcity of F stars for which we have measured
$\epsilon$ unambiguously. Scenario B appears to lie towards the top of any range we could expect for $\epsilon$ for a
star of its effective temperature. On the other hand, Scenario A appears to be at the minimum. Expecting a correction
for near-surface effects, we are inclined to believe Scenario B is more likely, but we cannot rule out the alternative.
We anticipate that a measurement of $\epsilon$ in many F stars observed by {\it Kepler} could help clearly define the
observational $\epsilon$--$T_\mathrm{eff}$ relation, clarifying the correct mode identification in Procyon.

\section{Gravity-mode period spacings in red giant stars}\label{Gperiod}

\begin{figure}
\epsscale{1.2}
\plotone{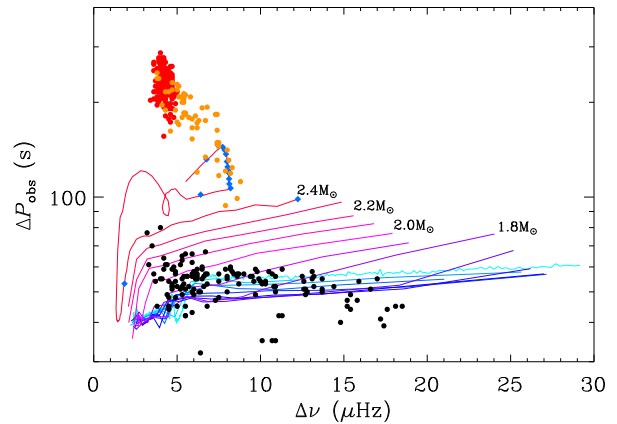}
\caption{Observed period spacing, $\Delta P_\mathrm{obs}$, as a function of the large separation $\Delta\nu$ in red
giants. Model tracks show the evolution of $\Delta P_\mathrm{obs}$ for hydrogen shell-burning red giants of near-solar
metallicity ($Z_0=0.017$) and masses between $1.0\,\mathrm{M}_\odot$ and $2.4\,\mathrm{M}_\odot$. The 
$2.4\,\mathrm{M}_\odot$ track extends past the tip of the RGB to helium-burning phases. Black dots show
$\Delta P_\mathrm{obs}$ as measured from {\it Kepler} red giant branch stars by \citet{Bedding11}. Red and orange dots
show red-clump and secondary-clump stars, as determined by \citet{Bedding11}. Blue diamonds are model calculations on
the $2.4\,\mathrm{M}_\odot$ track are equally spaced in time by 10\,Myr, starting from 516.5\,Myr. }\label{fig15}
\end{figure}

Theoretical studies of red giants predict a very dense frequency spectrum of non-radial modes
\citep{Dziembowski01, C-D04, Dupret09, Montalban10, DiMauro11, C-D11}. The vast majority are almost pure $g$ modes that are
largely confined to the core of the star and thus have surface amplitudes that are too low to be observable.
Asymptotically, these modes are expected to exhibit an approximately equal period spacing, which we denote 
$\Delta P_\mathrm{g}$ \citep{Tassoul80}.

There also exist $p$ modes in the convective envelope of the star that are equally spaced in frequency. As mentioned in the
Introduction, $p$ and $g$ modes of the same angular degree may undergo coupling. The modes undergoing this interaction take
on a mixed character, with $p$-mode characteristics in the envelope and $g$-mode characteristics in the core. The mixed
modes with enough $p$-mode character will be observable due to their reduced mode inertias and we can expect to see a
few $l=1$ modes for each $p$-mode order. Due to the weaker coupling between $l=2$ $p$ and $g$ modes, few additional
mixed modes are likely to be observed. Early {\it Kepler} results revealed that there were multiple $l=1$ peaks due to
mixed modes in each radial order \citep{Bedding10c}, which have recently been shown to have identifiable period spacings
\citep{Beck11}. Since these gravity-dominated mixed modes propagate deeply into the core of the star, they are key
indicators of core structure. \citet{Bedding11} found that the observed period spacing of the mixed $l=1$ modes can
distinguish red giants burning helium in their cores from those still only burning hydrogen in a shell. Similar results 
have also been found with {\it CoRoT} data \citep{Mosser11b}.

As mentioned, in the absence of any interaction with $p$ modes, the $g$ modes will be approximately equally spaced in
period. However, due to mode bumping, the observed period spacing ($\Delta P_\mathrm{obs}$) is substantially smaller
than the `true', asymptotic $g$-mode period spacing ($\Delta P_\mathrm{g}$). \citet{Bedding11} have shown that for red
giants with the best signal-to-noise ratio, in which several mixed modes are observed in each radial order, the true
period spacing may be recovered. However, for the vast majority of stars in their sample, this was not possible. Instead
they measured the average period spacing of the observed $l=1$ modes, $\Delta P_\mathrm{obs}$. We have done likewise 
with models.

To measure $\Delta P_\mathrm{obs}$ from models it is first necessary to determine which modes would be observable.
The mode with the lowest inertia in each order has the greatest amplitude and we calculated the period spacing
between this mode and the two adjacent modes. We take the average of these period spacings in the range
$\nu_{\mathrm{max}}\pm0.25\nu_{\mathrm{max}}$ to calculate $\Delta P_\mathrm{obs}$. In general, there will be variation in
the number of $l=1$ modes observable in each order and detailed modeling of any star will have to take this into
account. Nevertheless, our method of including only the central three modes is sufficient to follow the expected
evolution and mass dependence of $\Delta P_\mathrm{obs}$.

In Figure \ref{fig15} the solid lines show $\Delta P_\mathrm{obs}$ as a function of $\Delta\nu$ for our red-giant models.
Except for the highest mass model ($M=2.4\,\mathrm{M}_\odot$), we have not evolved the models past the helium ignition.
Stars on the RGB evolve from right to left in this figure and we see a gradual decrease in $\Delta P_\mathrm{obs}$ through
most of this phase. For stars with $M<1.8\,\mathrm{M}_\odot$ we see only a weak dependence on mass. We confirm
that $\Delta P_\mathrm{obs}$ as determined from the models for these lower-mass stars ($\sim 50$\,s) matches well with
the $\Delta P_\mathrm{obs}$ seen in observations of red giants \citep[$40 - 60$\,s;][]{Bedding11}, as shown in Figure
\ref{fig15}. Higher-mass stars show larger $\Delta P_\mathrm{obs}$. This raises the possibility that higher-mass RGB 
stars could be mistaken for helium-core-burning stars, although we note that few of these higher-mass stars
may be expected to be observed due both to their lower abundance relative to lower-mass stars and their rapid evolution.

Following the evolution of low-mass stars past the tip of the red giant branch is difficult because they undergo a
helium flash. Models computed from the helium main sequence by Montalb\'an et al. (in prep.) indicate period spacings
around $\sim 200$\,s, in agreement with observations of core helium-burning stars \citep{Bedding11}. 
We were able to compute the evolution of a higher-mass star ($2.4\,\mathrm{M}_\odot$), which undergoes
a more gradual onset of helium-core burning. After the model reaches the tip of the RGB, the period spacing increases 
rapidly due to the onset of a convective core \citep{C-D11}, followed by an increase in the large separation. The loop in 
the track during this phase occurs during a lull in the energy generation from the core; helium ignition occurs in 
pulses for this higher-mass model. The model then settles onto the secondary clump, where it
spends a (relatively) long period of time, with period spacing increasing slowly while the large separation remains
relatively constant. The model period spacings agree well with the observations by \citet{Bedding11}. As the model
begins to move up the asymptotic giant branch, the period spacing and large separation decrease once more. To help
indicate the evolution through this diagram, the blue diamonds in
Figure \ref{fig15} are equally spaced in age by 10\,Myr, starting from 516.5\,Myr. Clearly, the secondary-clump stage of
evolution is the longest, which is why we observe many more stars in this stage of their evolution than in other
stages \citep{Girardi99}. The agreement of this track with observations is excellent.

\section{Summary of Conclusions}\label{Conclude}

We have investigated the evolution of several measurements of the oscillation spectra of stars that
exhibit solar-like oscillations. We conclude:
\begin{enumerate}
\item The standard scaling relation for $\Delta\nu$ with density (equation~(\ref{scaling})) does quite well for lower-mass 
main-sequence stars, being correct to within a few percent, but larger deviations are found for other masses and 
evolutionary states. These deviations are predominantly a function of effective temperature and we suggest that the 
scaling relation may be improved by including a function of $T_\mathrm{eff}$ (equation~(\ref{scaling2})).
\item For main-sequence stars, their position in the \mbox{C-D} diagram is able to significantly constrain
their mass. Age is less well-constrained due to the uncertainty in $\delta\nu_{02}$ generally being
relatively larger than that in $\Delta\nu$. When stars evolve into subgiants the tracks converge, which limits the
ability of the relationship between $\Delta\nu$ and $\delta\nu_{02}$ to constrain the mass and age. 
\item In the $\epsilon$~diagram, the degeneracy of the evolutionary tracks during the subgiant stage is partially
broken. This suggests that the position of stars in this diagram could help constrain their masses and ages,
although difficulties in measuring $\epsilon$ and the near-surface offset presently limit this diagram's usefulness.
We find the near-surface offset is non-negligible for red giants, contrary to earlier claims.
\item Measuring $\epsilon$ shows promise for mode identification in stars where this is ambiguous due to short
mode lifetimes. In particular, $\epsilon$ is seen to depend mostly on effective temperature. Provided
$T_\mathrm{eff}$ is known, the correct $\epsilon$ and therefore the correct mode identification may be deduced.
\item We have investigated the evolution of g-mode period spacings ($\Delta P_\mathrm{obs}$) in models of red giant
stars. Models with masses below $\sim1.8\,\mathrm{M}_\odot$ show a weak dependence of $\Delta P_\mathrm{obs}$
on mass. For higher-mass stars, the observed position of the secondary clump in the $\Delta P_\mathrm{obs}$--$\Delta\nu$ diagram is well
matched by the models.
\end{enumerate}

\acknowledgments
We acknowledge the support of the Australian Research Council. TRW is supported by an Australian Postgraduate Award, a
University of Sydney Merit Award, an Australian Astronomical Observatory PhD Scholarship and a Denison Merit Award.

\end{document}